\documentclass[apj]{emulateapj}
\usepackage{amsmath}
\usepackage{graphicx}
\usepackage{amssymb}
\usepackage{times}

\begin{document}

\title{Maser and other instabilities in a weakly magnetized relativistic plasma:\\ Theory and the astrophysical relevance of the maser}

\author{Andrei Gruzinov}
\affiliation{Physics Dept., New York University,  726 Broadway, New York, NY 10003, USA}
\author{Eli Waxman}
\affiliation{Dept. of Particle Phys. \& Astrophys., Weizmann Institute of Science, Rehovot 76100, Israel}

\begin{abstract}

A sufficient condition for maser instability in a weakly magnetized relativistic plasma with an isotropic particle distribution function is given. The maser growth rates and polarizations are computed starting from the exact dielectric permittivity tensor of a magnetized plasma. For very weak magnetic fields, our results confirm the approximate validity of the 'standard maser theory', which is based on the Einstein coefficients method, with one significant exception. For inclined propagation and realistic (small but finite) field, the growth rates of the two (nearly circular) polarizations differ significantly, while the standard theory predicts two (nearly circular) polarizations with similar growth rates. We show that this deviation is due to circularly polarized synchrotron emission, which is neglected in the standard theory.

The maser is shown to grow {\it slower} than Langmuir waves. Nevertheless, significant generation of EM waves is seen in (highly simplified) direct numerical simulations. We study the nonlinear saturation of the maser instability and find that it offers a mechanism for the conversion of a significant fraction of the plasma energy into radio waves. We briefly discuss the conditions under which the maser instability may operate in astrophysical sources, and provide rough estimates that may be used as a guidance when studying particular astrophysical sources/phenomena.

\end{abstract}

\maketitle

\section{Introduction}
\label{sec:intro}

Collisionless shocks are responsible for many, if not most, of high-energy astrophysics phenomena. In particular, \cite{W17} proposed that Fast Radio Bursts (FRB) are emitted by a maser instability in a plasma created by a collisionless shock. \cite{W17} assumes that a collisionless shock creates a plasma with ultrarelativistic weakly magnetized electrons with a "hollow" distribution function (with particle density that increases with momentum, see \S~\ref{sec:results} for an exact definition). The maser instability is then calculated as in \cite{SW02}, following the general methodology of \cite{Ginzburg89} -- to which we refer as the 'standard maser theory'. The motivation for suggesting that collisionless shocks produce hollow distributions is based on the fact that plasma instabilities are expected to isotropize the momenta of the streaming particles faster than they facilitate full quasi-thermalization.

The maser of \cite{SW02} can be described by the standard maser theory only if the magnetic field is very weak. Even for a weak field, the standard theory is based on a number of additional approximations, most notably, assumptions regarding the polarization of the maser modes. Besides, the standard maser theory ignores the non-electromagnetic plasma modes, which may grow faster than the maser and change the plasma distribution function before the maser develops. The main purpose of our current paper is to check the validity of the standard maser theory by an exact brute-force computation of the plasma instabilities. The derivations, given in \S~\ref{sec:theory} and \S~\ref{sec:numerics}, are quite involved. We therefore open with a concise presentation of our findings in \S~\ref{sec:results}.

In \S~\ref{sec:relevance} we discuss the relevance of this type of maser in astrophysics, but already here we must address one obvious issue. As we will show, weakly magnetized ultra-relativistic electrons with an isotropic hollow distribution function do emit maser radio waves. However, we certainly do not know whether or not such hollow distributions are indeed created by collisionless shocks. At present, we do not have a complete theory of collisionless shocks, that will enable one to answer this question. In particular, numerical simulations \citep[e.g.][]{SironiKeshetLemoine15SSRv} describe only the very initial stages of the shock evolution. However, in a forthcoming paper we make an argument (based on looking at a large collection of numerical results) that relativistic collisionless shocks inevitably emit radio waves, at about the plasma frequency, at about 0.1--1\% efficiency. It might even be hard to say if the maser or other mechanisms (discussed in our forthcoming paper) are at work.

In  \S~\ref{sec:theory} we present the exact dispersion law describing instabilities of a relativistic magnetized plasma with an isotropic particle distribution function. Then, for a weakly magnetized ultrarelativistic plasma, we derive approximate analytic expressions for the growth rates of the maser and Langmuir modes.

In  \S~\ref{sec:numerics} we solve the exact dispersion law numerically, confirming our analytic approximations for the growth rates, but also finding Bernstein-like modes, which our analytics did not treat at all. Then we describe a simplified PIC simulation which shows how the unstable modes saturate.

A brief discussion of the main conclusions is given in \S~\ref{sec:discussion}.

\section{Maser in a weakly magnetized plasma: main results}
\label{sec:results}

\subsection{Maser growth rate}
\label{sec:results-rates}

In \S~\ref{sec:theory} and \S~\ref{sec:numerics} we study plasma instabilities in a relativistic weakly magnetized plasma with an isotropic particle distribution function, a large characteristic Lorentz factor of the electrons, and a small magnetic energy fraction:
\begin{equation}
\label{eq:assum}
\gamma _c\gg 1,~~~\xi_B\equiv \frac{B^2}{8\pi n\gamma _cmc^2}\ll 1.
\end{equation}
We find that unstable maser modes exist for a "hollow particle distribution" provided that $\gamma_c^2\xi_B>1$ (see \S~\ref{sec:crit}). By "hollow" we mean a distribution with momentum space particle density, $F$, that increases with momentum, i.e. $dF/d\gamma>0$, or $dn/d\gamma\propto \gamma^2 dF/d\gamma$ rising faster than $\gamma^2$.

The maser modes grow fastest (see fig.~\ref{fig:emgr}, \S~\ref{sec:growth}) near the modified Razin frequency,
\begin{equation}
\label{eq:wR}
\omega_R\equiv\Big(\frac{9}{2}\xi_B\Big)^{-1/4}\omega_p,
\end{equation}
where their growth rate is approximately
\begin{equation}
\label{eq:growth}
\omega_{IR}=\frac{1}{8\sqrt{3}}\left(\frac{\omega_B}{\omega_p}\right)^{3/2}\omega_p.
\end{equation}
Here, $\omega_p$ is the plasma frequency ($\omega_p^2\approx4\pi ne^2/(\gamma_c m)$, see eqs.~\ref{eq:Om} and~\ref{eq:omp0}), and $\omega_B=(3/2)eB/(\gamma_c m c)$. Note that $(\omega_B/\omega_p)^2=(9/2)\xi_B$.

The above results are similar to those obtained using the standard maser theory. Moreover, for mono-energetic electrons and perpendicular propagation, i.e. for wave vectors perpendicular to the magnetic field direction, the analytic expression derived for the growth rates from the exact dispersion relation, eq.~(\ref{eq:monem}), is identical to that obtained by the standard maser theory, eqs.~(34-37) of \citet{W17} (in the $g\gg1$ limit; note that the negative-absorption coefficient $\alpha_\nu$ is related to $\omega_I$ by $\alpha_\nu=2\omega_I/c$, with a factor of 2 due to the fact that $\alpha_\nu$ describes the evolution of the intensity while $\omega_I$ describes the evolution of the amplitude). This exact agreement supports the validity of both current and earlier analytic work, as well as the validity of our current numerical results, which agree with the analytic results in the appropriate limits.

\subsection{Polarization}
\label{sec:results-pol}

A significant deviation of our exact results from those of the approximate standard theory is related to the polarization of the growing modes. In order to rigorously determine the polarizations, and the growth rates, one needs to obtain the eigenmodes of the dispersion equation, $(c/\omega)^2(k^2E_i-k_ik_jE_j)=\epsilon_{ij}E_j$ where $\bf{E}$ is the electric field and $\epsilon$ is the full plasma permittivity. In the standard maser theory, this is avoided using the following line of arguments. In the absence of a magnetic field the plasma is isotropic, the dielectric tensor is diagonal, $\epsilon_{ij}=n^2\delta_{ij}$, and the refraction index is independent of polarization, $n^2=(ck/\omega)^2=1-(\omega_p/\omega)^2$. It is therefore assumed that in the limit $\xi_B\rightarrow0$, the effect of the plasma may be described simply by modifying $c$ to $c/n(\omega)$, and hence that the polarization of the modes is the same as the vacuum synchrotron modes, i.e. linear polarization parallel or perpendicular to the projection of $\mathbf{B}$ onto the plane perpendicular to $\mathbf{k}$. The growth rates for the two polarizations are then determined by calculating the (negative) absorption coefficients using the Einstein relations between emission and absorption, and synchrotron emission is calculated with $c$ replaced by $c/n$.

How small should $\xi_B$ be for the above approximation to hold? The magnetic field introduces anisotropy, with off-diagonal terms in the dielectric tensor. For a weak field, cold plasma and inclined propagation, with $\mathbf{k}$ not perpendicular to $\mathbf{B}$, the modes are circularly polarized with a frequency difference (between modes of a given $k$) of the order of the non-diagonal terms,
\begin{equation}\label{eq:dw}
  \frac{\Delta\omega}{\omega}\approx\frac{\omega_p^2\omega_B}{\omega^3}.
\end{equation}
For a relativistic plasma, Eq.~(\ref{eq:dw}) approximately describes the real part of the shift $\Delta\omega$ between the two circularly polarized modes \citep[see \S~\ref{sec:incl} and][]{SW02}, with the electron mass replaced by $\gamma_c m$ in the definition of $\omega_p$ and $\omega_B$ (as given in \S~\ref{sec:results-rates} above). In the standard maser theory, it is thus assumed that the polarizations are the synchrotron vacuum polarizations as long as the imaginary correction to $\omega$, $\omega_I$, is $\omega_I/\omega\gg\Delta\omega/\omega$, since in this case it is expected to dominate the modification of dielectric tensor. When this condition is not satisfied, it is assumed that the emitted power is converted to circularly polarized modes, i.e. that the growing modes are circularly polarized with equal growth rates given by the average of the growth rates calculated for the linear polarization (parallel and perpendicular) modes \citep{Ginzburg89}.

For inclined (non perpendicular) propagation, $\omega_{IR}/\Delta\omega(\omega=\omega_R) \propto (\omega_p/\omega_B)^{1/2}\propto\xi_B^{-1/4}$. Thus, in the limit $\xi_B\rightarrow0$ we expect $\omega_I/\Delta\omega\gg1$ and hence linearly polarized growing modes. However, in practice, for inclined propagation and realistic (small but finite) field, we expect $\omega_I/\Delta\omega<1$ and circular growing modes. Taking into account the $1/(8\sqrt{3})$ factor of Eq.~(\ref{eq:growth}) and the fact that this fast growth is obtained at $\omega\approx 0.5\omega_R$ \citep[ see Fig.~\ref{fig:emgr} and Fig.~1 of][]{W17}, we have $\omega_I/\Delta\omega\approx (0.5^2/(8\sqrt{3})(9\xi_B/2)^{-1/4}$ and
\begin{equation}\label{eq:xiBmax}
   \omega_I/\Delta\omega>1 \quad {\rm for}\quad \xi_B<3\times10^{-8}.
\end{equation}

The polarizations of the growing modes predicted by the standard maser theory are qualitatively consistent with our exact results. For $\xi_B=10^{-3}$ and inclined propagation we find that the growing modes are nearly circularly polarized. However, the growth rates of the two modes are not similar. We show in \S~\ref{sec:incl} that this deviation is due to circularly polarized synchrotron emission, which is neglected in the standard theory.

\subsection{Competition with other modes and saturation}
\label{sec:results-sat}

Our analysis, based on a direct solution for the zeros of the exact dielectric tensor, enables us to compare the growth rate of the maser mode to those of non-electromagnetic modes. We find that the growth rate of Langmuir (electro-static) waves is larger than that of the maser mode. The ratio of Langmuir and maser modes growth rates is larger for smaller $\xi_B$ ($\propto \xi_B^{-5/12}$). For $\xi_B=10^{-3}$, which is the smallest value expected in the downstream of a collisionless shock, the growth rate ratio is only a few. Thus, the maser mode growth is not expected to be suppressed. Moreover, we have investigated numerically in \S~\ref{sec:numerics} the nonlinear evolution of the instability for $\xi_B=10^{-3}$, and found that the growth of both the maser and Langmuir modes saturates at a similar energy density of the waves, which is a few percent of the initial (uniform) magnetic field energy density (the Langmuir modes reach saturation faster).

Our results imply that for an isotropic "hollow" distribution of electrons in a weakly magnetized plasma, a significant fraction of the initial uniform magnetic field energy may be converted to electromagnetic radiation at the modified Razin frequency.

\section{The astrophysical relevance of the maser effect }
\label{sec:relevance}

As noted in the introduction, collisionless shock waves are responsible for a wide range of high energy astrophysical phenomena. The analysis of this paper implies that if such shocks produce hollow distribution functions, then (i) the maser instability may play a significant role in the thermalization of the particle distribution, and (ii) a significant fraction of the shock energy may be radiated away as electromagnetic waves at frequencies somewhat higher than the plasma frequency. The time scale for the maser amplification of the waves, $1/\omega_{IR}\sim 5\xi_B^{-3/4}\omega_p^{-1}$, implies that sufficient time for wave amplification is generally expected in all astrophysical scenarios.

The maser instability offers a mechanism for the conversion of a significant fraction of plasma energy into radio waves. Such conversion is difficult to achieve in general, and it may lead to coherent radio emission with very large brightness temperatures. In what follows we provide some order of magnitude estimates for the conditions under which bright coherent maser radio emission may be expected. These estimates may be used as a guidance when studying particular astrophysical sources/phenomena. Two shock configurations are considered: a shock propagating into a plasma at rest, and a shock propagating within a relativistically expanding plasma.

\subsection{Shock waves driven into a plasma at rest}
\label{sec:rest}

Let us consider first a fast shock wave driven into an e-p plasma at rest, with density $n$. We assume that the shock velocity $\beta=v/c\gg\sqrt{m_e/m_p}$ such that the post-shock electrons are highly relativistic. Assuming that the post-shock electrons are close to equipartition, we may approximate $\gamma_c m_e\approx \gamma \beta^2 m_p$ where $\gamma$ is the shock Lorentz factor. Noting that the post-shock plasma (proper) density is approximately given by $4\gamma n$, the post shock plasma frequency is $\omega_p^2\approx 16\pi n e^2/\beta^2m_p$, and the observed frequency of the emitted waves is
\begin{equation}\label{eq:wR_piston}
  \nu_R=\gamma\frac{\omega_R}{2\pi} \approx
  0.2\frac{\gamma}{\beta}\left(\frac{\xi_B}{10^{-3}}\right)^{-1/4}
  \left(\frac{n}{10^{10}\rm cm^{-3}}\right)^{1/2} {\rm GHz}.
\end{equation}
For a shock driven into a e$^\pm$ plasma of density $n_\pm$, $\gamma_c\approx\gamma\gg1$ and we have
\begin{equation}\label{eq:wR_piston_epm}
  \nu_R\approx
  7\gamma\left(\frac{\xi_B}{10^{-3}}\right)^{-1/4}
  \left(\frac{n_\pm}{10^{10}\rm cm^{-3}}\right)^{1/2} {\rm GHz}.
\end{equation}

A combination of large density and/or large Lorentz factor is therefore required in order to produce observed emission in radio waves. Given the large relevant densities, one should consider the possible effects of free-free absorption. The shocked plasma is hot, and will therefore not produce much free-free absorption. However, the plasma lying ahead of the shock may be optically thick. The free-free absorption coefficient is approximately given by
\begin{equation}\label{eq:ff}
  \alpha^{\rm ff}\approx 10^{-6}\left(\frac{T}{\rm 1~keV}\right)^{-3/2}
  \left(\frac{n}{10^{10}\rm cm^{-3}}\right)^{2}\left(\frac{\nu}{1~\rm GHz}\right)^{-2}{\rm cm^{-1}},
\end{equation}
where $T$ is the upstream plasma temperature.

Thus, in addition to large densities and/or highly relativistic velocities, a small system size,
\begin{equation}\label{eq:max_r}
  r<10^6\gamma^4(T/1~{\rm keV})^{3/2}\,{\rm cm},
\end{equation}
is required in order to avoid free-free absorption. Sufficiently large densities and velocities may be obtained, for example, in supernova explosions surrounded by dense CSM. Typical parameters in such cases are $n\sim10^{10}{\rm cm^{-3}}$ and $\beta\sim1/30$ \citep[e.g.][]{Ofek10}. However, while radio wave emission may be generated, the CSM extent is large, $\sim10^{14}$~cm, and the free-free absorption optical depth is large.

Possible sites for the generation of observable maser radio emission may be the hot corona regions around accretion disks near compact objects. The plasma density may be large enough in such systems for a (mildly) relativistic shock to produce radio maser emission, and the systems may be compact and hot enough to avoid free-free absorption.

\subsection{Shock waves within a relativistically expanding plasma}
\label{sec:jets}

Many high energy astrophysics phenomena are believed to be associated with shock waves propagating within a relativistically expanding plasma, often in the form of a jet. Let us therefore consider an e-p plasma jet of opening angle $\theta$ and expansion Lorentz factor $\gamma$. We do not discuss jets in which the energy flux is carried mainly by the magnetic field, since while maser emission may occur in a e$^\pm$ plasma, $\xi_B\ll1$ may not be satisfied in such jets.

The proper density of the plasma is $n(r)=(2/\theta^2)L_k/(4\pi r^2\gamma^2 m_p c^3)$ where $L_k$ is the kinetic luminosity of the jet (assuming a double sided jet and $\theta<<1$). A mildly relativistic shock traveling within the jet would heat electrons to $\gamma_c \approx m_p/m_e$. Expressing $r$ in terms of the observed signal duration, $r=2\gamma^2 c\Delta t$, we have
\begin{equation}\label{eq:wR_jet}
  \nu_R\approx
  0.1\frac{1}{\gamma^2\theta}\left(\frac{\xi_B}{10^{-3}}\right)^{-1/4}\frac{10^3\rm s}{\Delta t}
  \left(\frac{L_K}{10^{46}\rm erg/s}\right)^{1/2} {\rm GHz}.
\end{equation}
Assuming that the jet is driven by a black hole of mass $M$ with $\Delta t=R_s/c=2 GM/c^3$, we may write this as
\begin{equation}\label{eq:wR_jet_M}
  \nu_R\approx
  0.1\frac{1}{\gamma^2\theta}\left(\frac{\xi_B}{10^{-3}}\right)^{-1/4}
  \left(\frac{10^8M_\odot}{M}\right)^{1/2}
  \left(\frac{L_K}{L_E}\right)^{1/2} {\rm GHz},
\end{equation}
where the Eddington luminosity is $L_E=1.3\times10^{38}{M/M_\odot}{\rm erg/s}$.

Eq.~(\ref{eq:wR_jet}) may also be used to estimate the maser emission frequency in the case of the deceleration of a relativistic expanding shell, that was emitted impulsively (over time $<\Delta t$). The reverse shock driven into the shell, as it is being decelerated by an external medium, is mildly relativistic when it crosses most of the decelerating shell \citep[i.e. when a significant fraction of the kinetic energy is converted to thermal energy and a significant fraction of the energy is transferred to the external medium, e.g.][]{W17}. The maser emission frequency of the reverse shock is obtained from the above equations with $\Delta t$ taken as the observed pulse duration and $L_K$ given by the shell's energy $E$ via $L_K=E/\Delta t$. This yields, for example, the expected maser frequency for FRBs as obtained in \cite{W17}.

Here too, free-free absorption may suppress the emission of radio waves. The shock-heated electrons will cool via synchrotron emission to a Lorenz factor $\gamma_m$ at which their (proper) energy loss time, $(m_ec)/(4\sigma_T \gamma_m U_B/3)$ is comparable to the (proper) expansion time, $r/\gamma c$, $\gamma_m=(3\pi)(\gamma^3/\xi_B)(m_e c^3 r)/(\sigma_T L)$. Replacing $\Delta t=2GM/c^3$, we find
$\gamma_m=400(\gamma/3)^5(10^3\xi_B)^{-1}(L/L_E)^{-1}$. That is, for $\gamma=$ a few and $L\approx L_E$ the electrons cool to $T\approx 100$~MeV. The free-free absorption optical depth of the cooled electrons at $\nu_R$, $\tau^{\rm ff}_R\approx\alpha^{\rm ff}(\nu_R) r/\gamma$, is
\begin{equation}\label{eq:wR_jet_ff}
  \tau^{\rm ff}_R\approx 0.3 \left(\frac{\xi_B}{10^{-3}}\right)^{2}
 \left(\frac{3}{\gamma}\right)^{\frac{29}{2}}\left(\frac{0.1}{\theta}\right)^{5}
  \left(\frac{10^3\rm s}{\Delta t}\frac{L_K}{10^{46}\rm erg/s}\right)^{5/2},
\end{equation}
or, for $\Delta t=2GM/c^3$,
\begin{equation}\label{eq:wR_jet_ff_M}
  \tau^{\rm ff}_R\approx 0.5 \left(\frac{\xi_B}{10^{-3}}\right)^{2}
 \left(\frac{3}{\gamma}\right)^{\frac{29}{2}}\left(\frac{0.1}{\theta}\right)^{5}
  \left(\frac{L_K}{L_E}\right)^{5/2}.
\end{equation}

The above equations show that coherent maser emission may be produced by relativistic winds driven by black-holes over a wide range of masses, provided that the energy flux is dominated by kinetic energy at the dissipation region.

\section{Theory of the maser instability}
\label{sec:theory}

Even under the simplifying assumptions of eq.~(\ref{eq:assum}), highly relativistic electrons and weak magnetization, the problem of calculating the maser growth rates and polarizations remains intractable analytically. In the next section we solve the problem numerically.

It turns out, however, that a physically interesting case of tangled background magnetic field does admit an analytic treatment. For a tangled magnetic field, the radiation is unpolarized, and the maser is fully characterized by the polarization-averaged growth rate, to be computed in \S~\ref{sec:growth}.

Before computing the averaged growth rate, we must do a lot of other analytic work.  In the following subsections we:
\begin{enumerate}

\item Write down the exact dispersion relation. This is needed for the numerical work of the next section and for the analytic work of this section.

\item Write down and solve a simplified dispersion relation, describing high-frequency EM and Langmuir waves in unmagnetized relativistic plasma. This is needed as a check of the numerics of the next section and for the analytic work of this section.

\item Calculate the growth rates for perpendicular propagation. This is needed to show that our results are in exact agreement with some of the results of \cite{SW02, W17}, which these authors obtained by the standard maser theory.  The exact agreement gives credence to analytic computations of both ours and the previous papers, and also to our numerics, which agrees with the analytic results where applicable.

The perpendicular propagation case is also needed to show that Langmuir waves are unstable, to establish the growth rate scalings,  and for further analytic work of this section.

\item Establish a sufficient condition for the maser instability.

\item Compute the polarization-averaged growth rate for inclined propagation.

\item Explain the different growth rates of two nearly circular polarizations.

\end{enumerate}

\subsection{Permittivity tensor}
\label{sec:perm}

From the plasma physics perspective, the maser effect is just one of the many possible instabilities of magnetized collisionless plasma. To calculate the eigenmodes and their growth rates, one proceeds along standard lines. Take the EM field and the distribution function perturbations $\propto e^{-i\omega t+i{\bf k}\cdot {\bf r}}$. The eigenmode equation for the electric field is
\begin{equation}\label{eq:pol}
\tilde{\epsilon}_{ij}E_j=0,
\end{equation}
and the corresponding dispersion relation is
\begin{equation}\label{eq:disp}
\det \tilde{\epsilon} =0.
\end{equation}
Here
\begin{equation}\label{eq:tild}
\tilde{\epsilon}_{ij}\equiv \epsilon_{ij}+\frac{k_ik_j}{\omega^2}-\frac{k^2}{\omega^2}\delta_{ij},
\end{equation}
the speed of light $c=1$, and $\epsilon$ is the plasma dielectric permittivity tensor, given by, e.g. \cite{Aleksandrov84}
\begin{equation}\label{eq:eps}
\epsilon_{ij}=\delta_{ij}+\frac{\Omega _p^2}{\omega}\int d^3p\frac{dF}{d\gamma}\sum_n\frac{\Pi^{(n)}_{ij}}{\omega -k_zv_z-n\Omega_B/\gamma},
\end{equation}
\begin{equation}\label{eq:Om}
\Omega _p^2\equiv \frac{4\pi ne^2}{m},~~~\Omega_B\equiv \frac{eB}{m}.
\end{equation}
In Eq.~(\ref{eq:Om}), $n$ is the number density. In Eq.~(\ref{eq:eps}), $F$ is the distribution function, which is assumed to be isotropic, and normalized by
\begin{equation}
\int d^3pF=1.
\end{equation}
The background magnetic field $B$ is along $\hat{z}$. The wavevector is ${\bf k}=(k_x,0,k_z)$. The velocity components $v_z$ and $v_\perp$ are along and perpendicular to $\hat{z}$, $\gamma$ is the Lorentz factor of the particle.
The sum in Eq.~(\ref{eq:eps}) is over all integers $n$, and the $n$-dependent tensor is

\begin{equation}
\Pi^{(n)}\equiv \left(
\begin{array}{ccc}
v_\perp^2\frac{n^2}{b^2}J_n^2 & iv_\perp^2\frac{n}{b}J_nJ_n' & v_\perp v_z\frac{n}{b}J_n^2  \\

\\

-iv_\perp^2\frac{n}{b}J_nJ_n' & v_\perp^2J_n'^2  & -iv_\perp v_zJ_nJ_n' \\

\\

v_\perp v_z\frac{n}{b}J_n^2&  iv_\perp v_zJ_nJ_n' & v_z^2J_n^2
\end{array}
\right),
\end{equation}
where the Bessel functions and their derivatives are
\begin{equation}
J_n\equiv J_n(b),~~J_n'\equiv J_n'(b),~~b\equiv \frac{k_xv_\perp \gamma}{\Omega_B}.
\end{equation}

For actual calculations, we rename the permittivity components:
\begin{equation}
\epsilon\equiv\left(
\begin{array}{ccc}
\epsilon_1 & ig_1 & g_2  \\

\\

-ig_1 & \epsilon_2  & -ig_3 \\

\\

g_2 &  ig_3 & \epsilon_3
\end{array}
\right),
\end{equation}
and write the dispersion relation as
\begin{equation}\label{eq:edisp}
\tilde{\epsilon}_1\tilde{\epsilon}_2\tilde{\epsilon}_3 - g_1^2\tilde{\epsilon}_3 - \tilde{g}_2^2\tilde{\epsilon}_2 - g_3^2\tilde{\epsilon}_1+2g_1\tilde{g}_2g_3 = 0,
\end{equation}
where the tilde has the same meaning as before:
\begin{equation}
\begin{array}{c}
\tilde{\epsilon}_1\equiv \epsilon_1-\frac{k_z^2}{\omega^2},~~~\tilde{\epsilon}_3\equiv \epsilon_3-\frac{k_x^2}{\omega^2},
\\
\\
\tilde{\epsilon}_2\equiv \epsilon_2-\frac{k^2}{\omega^2},~~~ \tilde{g}_2\equiv g_2+\frac{k_xk_z}{\omega^2}.
\end{array}
\end{equation}

\subsection{Permittivity and eigenmodes at zero magnetization}
\label{sec:unmag}

For $B=0$, standard calculations give
\begin{equation}\label{eq:eps0}
\epsilon_{ij}=\delta_{ij}+\frac{\Omega _p^2}{\omega}\int d^3p\frac{dF}{d\gamma}\frac{v_iv_j}{\omega -{\bf k}\cdot {\bf v}}.
\end{equation}
From Eq.~(\ref{eq:eps0}), one gets two different permittivities for transverse and longitudinal polarizations:
\begin{equation}\label{eq:epsd}
\epsilon_{ij}=\epsilon_\perp(\delta_{ij}-\hat{k}_i\hat{k}_j)+\epsilon_\parallel\hat{k}_i\hat{k}_j,
\end{equation}
\begin{equation}\label{eq:epsp}
\epsilon_\perp=1+\frac{\Omega _p^2}{2\omega}\int d^3p\frac{dF}{d\gamma}v^2\frac{1-\mu^2}{\omega -kv\mu},
\end{equation}
\begin{equation}\label{eq:epspa}
\epsilon_\parallel=1+\frac{\Omega _p^2}{\omega}\int d^3p\frac{dF}{d\gamma}v^2\frac{\mu^2}{\omega -kv\mu},
\end{equation}
\begin{equation}
\mu \equiv \cos (\hat{k}\cdot{\bf v}).
\end{equation}

The dispersion relations for transverse (EM waves) and longitudinal (Langmuir waves) polarizations are
\begin{equation}\label{eq:disp00}
\omega ^2=\epsilon_\perp k^2,~~~\epsilon_\parallel=0.
\end{equation}
We are interested in the high-frequency, meaning $k\gg \gamma _c^{-1/2} \Omega _p$, EM waves. As we are about to show, for these large wavenumbers, Eq.~(\ref{eq:disp00}) gives $\omega \approx k$, allowing to simplify the fraction in Eq.~(\ref{eq:epsp}):
\begin{equation}
\frac{1-\mu^2}{\omega-kv\mu}\approx \frac{1+\mu}{\omega },
\end{equation}
where we have used $v\approx 1$, for $\gamma _c\gg 1$. Now Eq.~(\ref{eq:epsp}) gives
\begin{equation}\label{eq:epsp0}
\epsilon_\perp\approx 1-\frac{\omega _p^2}{\omega^2},
\end{equation}
\begin{equation}\label{eq:omp0}
\omega _p^2\equiv \Omega _p^2\int d^3pF\frac{1}{\gamma}.
\end{equation}
\newline
Recalling the normalization, $\int d^3pF=1$, we can write Eq.~(\ref{eq:omp0}) in a clearer form
\begin{equation}\label{eq:omp}
\omega _p^2\equiv \Omega _p^2<\frac{1}{\gamma}>\sim \frac{4\pi ne^2}{\gamma _cm}.
\end{equation}

Now Eqs.~(\ref{eq:disp}, \ref{eq:tild}, \ref{eq:epsd}, \ref{eq:epsp0}) give the standard plasma dispersion law
\begin{equation}\label{eq:om0}
\omega^2=k^2+\omega _p^2.
\end{equation}
The numerical calculations of \S\ref{sec:numerics}, although based on the exact permittivity, Eq.~(\ref{eq:eps}), do confirm that the frequency $\omega$ is approximately given by Eq.~(\ref{eq:om0}) at $k\gg \omega_p$.

We are also interested in high-frequency Langmuir waves, with $\omega\approx k\gg \omega_p$ (in practice, $\frac{\omega}{\omega_p}\sim$few). Such waves exist only in ultra-relativistic plasma, where we can put $v=1$ in Eq.~(\ref{eq:epspa}) and calculate the $\mu$-integral as
\begin{equation}
\int\limits_{-1}^{1}d\mu \frac{\mu^2}{\omega -k\mu}\approx \frac{1}{k}\ln \frac{\omega}{\omega-k},
\end{equation}
giving the approximate longitudinal permittivity
\begin{equation}\label{eq:epa}
\epsilon_\parallel\approx 1-\frac{\omega _p^2}{\omega k}\ln \frac{\omega}{\omega-k}\approx 1-\frac{\omega _p^2}{ k^2}\ln \frac{k}{\omega-k},
\end{equation}
and the dispersion law
\begin{equation}
\omega \approx k\big( 1+e^{-\frac{k^2}{\omega_p^2}}\big).
\end{equation}

\subsection{Perpendicular propagation}
\label{sec:prop}
Taking $k_z=0$, i.e. a wave propagating perpendicular to the background magnetic field, we have $g_2=g_3=0$, $\tilde{g}_2=0$, and the dispersion law, Eq.~(\ref{eq:edisp}), splits into
\begin{equation}\label{eq:dipp}
\epsilon_1\tilde{\epsilon} _2-g_1^2=0,~~~\tilde{\epsilon} _3=0.
\end{equation}
This means that pure transverse modes are polarized along the magnetic field. Transverse modes with polarization perpendicular to the magnetic field get mixed with the longitudinal modes. However, for small $B$, the dispersion law $\epsilon_1\tilde{\epsilon} _2=g_1^2$ describes one mostly-transverse and one mostly-longitudinal mode, which we will call perpendicular-polarized EM and Langmuir waves.

We calculate the growth rates of the EM waves and the Langmuir waves in turn. Then, to illustrate the results, we compute the growth rates for a monoenergetic distribution.

\subsubsection{Perpendicular propagation: EM waves}\label{seq:emp}
The dispersion laws for the perpendicular and parallel polarizations of EM waves with perpendicular propagation are
\begin{equation}\label{eq:dispp}
\epsilon _2-\frac{k^2}{\omega^2}=\frac{g_1^2}{\epsilon_1}, ~~~\epsilon _3-\frac{k^2}{\omega^2}=0
\end{equation}
For zero magnetization, at high frequencies, $k\gg \omega_p$, we have
\begin{equation}\label{eq:perma}
\epsilon _2= \epsilon _3= \epsilon_\perp\approx 1-\frac{\omega _p^2}{\omega^2},~~~g_1= 0,
\end{equation}
and we get the same dispersion law for both polarizations:
\begin{equation}\label{eq:freq}
\omega =\sqrt{k^2+\omega _p^2}\approx k+\frac{\omega_p^2}{2k}.
\end{equation}

At non-zero but small magnetization, $\xi_B \ll 1$, there will be corrections to the permittivities Eq.~(\ref{eq:perma}), and, correspondingly, to the frequencies of the two modes. To leading order, the corrected frequencies are given by Eq.~(\ref{eq:dispp}):
\begin{equation}\label{eq:dispp2}
1-\frac{k^2+\omega _p^2}{\omega^2}=-\delta \epsilon _2+\frac{g_1^2}{\epsilon_1},
\end{equation}
\begin{equation}\label{eq:dispp1}
1-\frac{k^2+\omega _p^2}{\omega^2}=-\delta \epsilon_3.
\end{equation}

We are interested in the maser growth rates, which are the imaginary parts of the frequency corrections. From Eqs.~(\ref{eq:dispp1}, \ref{eq:dispp2}), we get the approximate growth rates
\begin{equation}\label{eq:rates0}
\omega _{I\perp}\approx -\frac{1}{2}{\rm Im} (\epsilon_2)\omega , ~~~\omega _{I\parallel}\approx -\frac{1}{2}{\rm Im} (\epsilon_3)\omega.
\end{equation}
Here $\omega$ is the real part of the frequency, the parallel and perpendicular polarizations are with respect to the background magnetic field, and $|{\rm Im }(\frac{g_1^2}{\epsilon_1})|\ll |{\rm Im} (\epsilon_2)|$ will be shown below.

The imaginary parts of the permittivities are calculated at real frequency, Eq.~(\ref{eq:freq}), by Landau's replacement,
\begin{equation}
\frac{1}{\omega +...}\rightarrow \frac{1}{\omega +i0 +...}={\cal P}\frac{1}{\omega +...}-i\pi \delta (\omega +...).
\end{equation}
Consider first ${\rm Im} (\epsilon_2)$. The starting expression is
\begin{equation}
{\rm Im} (\epsilon_2)= -\pi \frac{\Omega _p^2}{\omega}\int d^3p\frac{dF}{d\gamma}v_\perp ^2\sum_nJ_n'^2\delta(\omega -n\Omega_B/\gamma ).
\end{equation}
For small $B$, we replace the sum over $n$ by the integral over $n$, and calculate the integral over $n$ by removing the delta-function:
\begin{equation}\label{eq:ime1}
{\rm Im} (\epsilon_2)= -\pi \frac{\Omega _p^2}{\Omega _B\omega}\int d^3p\frac{dF}{d\gamma}\gamma v_\perp ^2J_n'^2,~~~n\equiv \frac{\gamma \omega}{\Omega _B}.
\end{equation}

Since $n$ is large, we can approximate the Bessel function by the Airy function (here and below, our statements regarding special functions come from \cite{AS}):
\begin{equation}\label{eq:besa}
J_n'(b)\approx -\Big( \frac{2}{n}\Big) ^{2/3}{\rm Ai}'\Big[ \Big( \frac{n}{2}\Big) ^{2/3}\Big(1-\frac{b^2}{n^2}\Big)\Big].
\end{equation}
Here $b=\frac{kv_\perp \gamma}{\Omega_B}$, and
\begin{equation}\label{eq:bon}
\frac{b}{n}=\frac{kv_\perp}{\omega}\approx1-\frac{\omega _p^2}{2\omega^2}-\frac{\psi^2}{2},
\end{equation}
where $\psi$ is the angle between the particle velocity and the $xy$-plane, which is assumed to be small (to be justified momentarily). We have also used the approximate dispersion law Eq.~(\ref{eq:freq}), and we put $v=1$, corresponding to $\gamma _c>>1$.

With the replacements of Eqs.~(\ref{eq:besa},\ref{eq:bon}), Eq.~(\ref{eq:ime1}) takes the form
\begin{equation}\label{eq:ime2}
\begin{array}{c}
{\rm Im} (\epsilon_2)= -\pi \frac{\Omega _p^2}{\Omega _B\omega}\int\limits_0^\infty 2\pi p^2dp\frac{dF}{d\gamma}\gamma
\\

\\
\times \Big(\frac{2\Omega _B}{\gamma \omega}\Big)^{4/3}\int\limits_{-\infty}^\infty d\psi{\rm Ai}'^2\Big[ \Big( \frac{\gamma \omega}{2\Omega _B}\Big) ^{2/3}\Big(\frac{\omega_p^2}{\omega^2}+\psi^2\Big)\Big].
\\

\\
\end{array}
\end{equation}
We now see that only $\psi \ll 1$ contribute to the integral, because the Airy function decreases exponentially at large positive argument.

The $\psi$-integral in Eq.~(\ref{eq:ime2}) can be replaced by a more familiar expression, used in the theory of synchrotron radiation. With the help of the identity
\begin{equation}
\begin{array}{c}
\int\limits_{-\infty}^\infty d\psi{\rm Ai}'^2\big[ x^{2/3}( 1+\psi^2)\big]=\frac{x^{1/3}}{4\sqrt{3}\pi}f_\perp\big(\frac{4}{3}x\big),
\\

\\
f_\perp(x)\equiv \int\limits_x^\infty dyK_{5/3}(y)+K_{2/3}(x),
\end{array}
\end{equation}
we re-write Eq.~(\ref{eq:ime2}) as
\begin{equation}\label{eq:ime3}
\begin{array}{c}
{\rm Im} (\epsilon_2)= -\frac{1}{4\sqrt{3}}\frac{\omega _p^2\Omega _p^2}{\omega^4}\int d^3p\frac{dF}{d\gamma}f_\perp(x),
\\

\\
x\equiv\frac{2}{3}\frac{\gamma \omega_p^3}{\Omega_B\omega^2}.
\\

\\
\end{array}
\end{equation}
We have checked, by numerically computing the exact permittivity Eq.~(\ref{eq:eps}), that Eq.~(\ref{eq:ime3}) is correct in the limit $\gamma _c\rightarrow \infty$, $\xi_B\rightarrow 0$, $\omega \sim \xi_B^{-1/4}\omega _p$, that is, for the frequencies where the maser operates.

Now, using Eq.~(\ref{eq:rates0}), we get the growth rate for perpendicular propagation with perpendicular polarization:
\begin{equation}\label{eq:ratep}
\begin{array}{c}
\omega _{I\perp}\approx \frac{1}{8\sqrt{3}}\frac{\omega _p^2\Omega _p^2}{\omega^3}\int d^3p\frac{dF}{d\gamma}f_\perp(x),
\\

\\
x\equiv\frac{2}{3}\frac{\gamma \omega_p^3}{\Omega_B\omega^2},
\\

\\
f_\perp(x)\equiv \int\limits_x^\infty dyK_{5/3}(y)+K_{2/3}(x).

\end{array}
\end{equation}
For perpendicular propagation with parallel polarization, we compute ${\rm Im}(\epsilon_3)$. The result is

\begin{equation}\label{eq:ratepa}
\begin{array}{c}
\omega _{I\parallel}\approx \frac{1}{8\sqrt{3}}\frac{\omega _p^2\Omega _p^2}{\omega^3}\int d^3p\frac{dF}{d\gamma}f_\parallel (x),
\\

\\
x\equiv\frac{2}{3}\frac{\gamma \omega_p^3}{\Omega_B\omega^2},
\\

\\
f_\parallel(x)\equiv  \int\limits_x^\infty dyK_{5/3}(y)-K_{2/3}(x).
\end{array}
\end{equation}
We have checked that our numerically computed growth rates, for perpendicular propagation, approximately agree with the analytic expressions Eqs.~(\ref{eq:ratep}, \ref{eq:ratepa}).

({\footnotesize  It remains to show that $|{\rm Im }(\frac{g_1^2}{\epsilon_1})|\ll |{\rm Im} (\epsilon_2)|$. Assuming $\omega \sim \xi_B^{-1/4}\omega _p$, we estimate, from Eq.(\ref{eq:ime2}),
\begin{equation}
{\rm Im} (\epsilon_2)\sim \xi_B.
\end{equation}
A similar calculation gives \begin{equation}
{\rm Im} (g_1)\sim \xi_B^{3/4}.
\end{equation}
Also, since $\epsilon_1\approx 1$ one expects, and can confirm numerically, that the dominant part of $g_1$, which is real, does not exceed the  leading part of the plasma dispersion: $|g_1|<\frac{\omega_p^2}{\omega^2}\sim \xi_B^{1/2}$. We can now estimate
\begin{equation}
|{\rm Im }(\frac{g_1^2}{\epsilon_1})|\sim |g_1{\rm Im} (g_1)|<\xi_B^{5/4}\ll \xi_B\sim |{\rm Im} (\epsilon_2)|,
\end{equation}
as stated and used above.
})

\subsubsection{Perpendicular propagation: Langmuir waves}

The dispersion relation is still the first of Eqs.~(\ref{eq:dipp}), which we now write as
\begin{equation}\label{eq:displ1}
\epsilon_1=\frac{g_1^2}{\tilde{\epsilon}_2}.
\end{equation}
Ignoring logarithmic factors, near the instability domain $\omega \sim k \sim \omega_p$, we have the following estimates
\begin{equation}
|{\rm Im}~ \epsilon_1| \sim |{\rm Re}~ \epsilon_1| \sim |\tilde{\epsilon}_2| \sim 1 \gg |g_1|,
\end{equation}
and we can approximate the dispersion relation Eq.~(\ref{eq:displ1}) simply as
\begin{equation}
\epsilon_1=0.
\end{equation}
When logarithmic factors are included, one gets $|{\rm Im}~ \epsilon_1| \ll |{\rm Re}~ \epsilon_1|$, allowing, as in \S \ref{seq:emp}, to approximate $\epsilon _1$ by its unmagnetized value plus the imaginary part. Using Eq.~(\ref{eq:epa}), we have
\begin{equation}
\epsilon_1\approx 1-\frac{\omega_p^2}{k^2}\ln \frac{k}{\omega -k}+i~{\rm Im}(\epsilon_1).
\end{equation}
This gives the growth rate
\begin{equation}
\omega _{LI}\approx -\frac{\omega^3}{\omega_p^2}e^{-\frac{k^2}{\omega_p^2}}~{\rm Im}(\epsilon_1).
\end{equation}
${\rm Im}(\epsilon_1)$ is calculated as in \S \ref{seq:emp}. The result is
\begin{equation}
\begin{array}{c}
{\rm Im} (\epsilon_1)= -\frac{1}{2\sqrt{3}}\frac{\Omega _p^2}{\omega^2}\int d^3p\frac{dF}{d\gamma}f_L(x),
\\

\\
f_L(x)\equiv  -\int\limits_x^\infty dyK_{5/3}(y)+2K_{2/3}(x),
\\

\\
x\equiv\frac{2^{5/2}}{3}\frac{\gamma \omega}{\Omega_B}e^{-\frac{3}{2}\frac{\omega^2}{\omega_p^2}},
\\

\\
\omega _{LI}\approx \frac{1}{2\sqrt{3}}\frac{\Omega _p^2\omega}{\omega_p^2}e^{-\frac{\omega^2}{\omega_p^2}}\int d^3p\frac{dF}{d\gamma}f_L(x).

\end{array}
\end{equation}

\begin{figure}
 \includegraphics[width=0.5\textwidth]{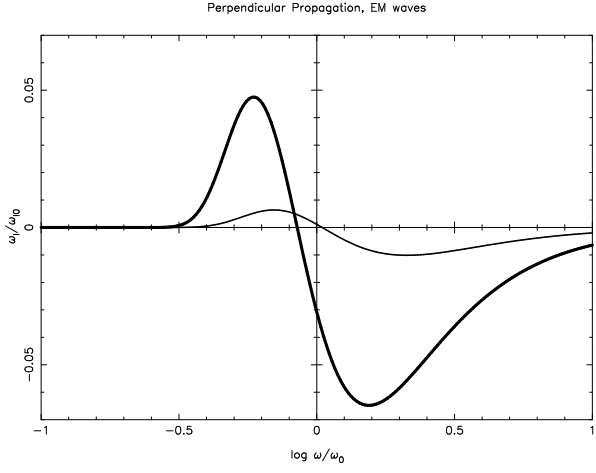}
\caption{EM growth rates, Eqs.~(\ref{eq:monem}), for perpendicular (thick) and parallel (thin) polarizations. The frequency $\omega$ is normalized to $\omega_0\equiv \Big(\frac{\omega_p}{\omega_B}\Big)^\frac{1}{2}\omega_p$, the growth rate $\omega_I$ is normalized to  $\omega_{I0}\equiv \Big(\frac{\omega_B}{\omega_p}\Big)^\frac{3}{2}\omega_p$.   }
\label{fig:emgr}
\end{figure}

\subsubsection{Perpendicular propagation: A monoenergetic distribution}\label{sec:mono}
\begin{figure}
 \includegraphics[width=0.5\textwidth]{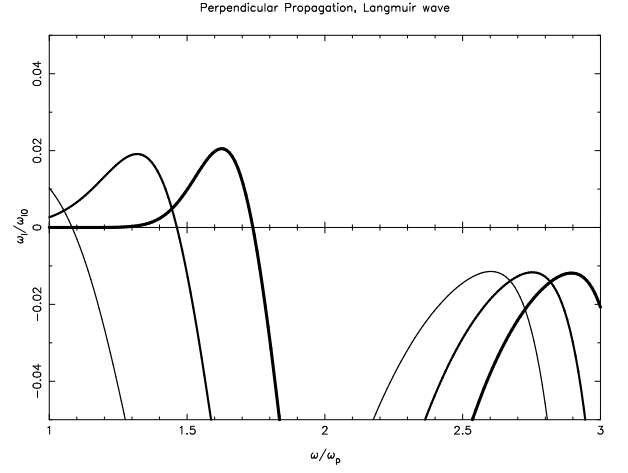}
\caption{Langmuir wave growth rates, Eq.~(\ref{eq:monl}), for $\xi_B=10^{-2}$ (thin),  $\xi_B=10^{-3}$ (thicker), $\xi_B=10^{-4 }$ (thick), and parallel polarizations. The frequency $\omega$ is normalized to  $\omega_p$, the growth rate $\omega_I$ is normalized to $\omega_{I0}\equiv \Big(\frac{\omega_B}{\omega_p}\Big)^\frac{2}{3}\omega_p$. The graph starts at $\omega =\omega_p$ because Eq.~(\ref{eq:monl}) is invalid at smaller $\omega$. }
\label{fig:lgr}
\end{figure}

For a monoenergetic  distribution,  the integrals in the growth rate expressions  are
\begin{equation}
\int d^3p\frac{dF}{d\gamma}f(x)=-\frac{1}{\gamma}(2f+xf'),
\end{equation}
giving
\begin{equation}\label{eq:monem}
\begin{array}{c}
\omega _{I\perp ,\parallel}=-\frac{1}{8\sqrt{3}}\Big(\frac{\omega_B}{\omega_p}\Big)^\frac{3}{2}\omega_px^\frac{3}{2}\Big(2f_{\perp ,\parallel}(x)+xf'_{\perp ,\parallel}(x)\Big),
\\
\\
x\equiv\frac{\omega_p^3}{\omega_B\omega^2}, ~~~\omega_B\equiv \frac{3}{2}\frac{\Omega_B}{\gamma},
\end{array}
\end{equation}
shown in Fig.~(\ref{fig:emgr}), and
\begin{equation}\label{eq:monl}
\begin{array}{c}
\omega _{IL}=-\frac{3^{1/6}}{2^{8/3}}\omega_B^\frac{2}{3}\omega^\frac{1}{3}x^\frac{2}{3}\Big(2f_L(x)+xf'_L(x)\Big),
\\
\\
x\equiv\frac{2^{5/2}}{3}\frac{\omega}{\omega_B}e^{-\frac{3}{2}\frac{\omega^2}{\omega_p^2}},
\end{array}
\end{equation}
shown in Fig.~(\ref{fig:lgr}).

\subsection{The maser instability criterion}
\label{sec:crit}
A simple sufficient condition for the maser instability can be established, which qualitatively means that hollow isotropic distributions are maser unstable. The exact sufficient condition is
\begin{equation}\label{eq:crit}
\frac{dF}{d\gamma}|_{\gamma =\sqrt{\frac{3}{2}}\frac{\Omega_B}{\omega_p}}>0.
\end{equation}
This condition makes sense only if $\gamma =\sqrt{\frac{3}{2}}\frac{\Omega_B}{\omega_p}>1$, or
\begin{equation}\label{eq:critc}
\gamma_c^2\xi_B\gtrsim 1.
\end{equation}
The inequality Eq.~(\ref{eq:critc}) is likely satisfied in the downstream of a relativistic shock in electron-proton plasma, because the equipartition values, $\gamma _c\gtrsim 1000$, $\xi_B\sim 1$, fulfill the inequality by a large margin.

To prove that Eq.~(\ref{eq:crit}) is a sufficient condition for maser instability, consider perpendicular propagation with parallel polarization. The exact dispersion law is
\begin{equation}
\tilde{\epsilon}_3=0.
\end{equation}
In the long-wavelength limit, $k\rightarrow 0$, the real part of the dispersion law can be calculated by setting $k=0$ in Eq.~(\ref{eq:eps}):
\begin{equation}
{\rm Re}(\tilde{\epsilon}_3)\approx \epsilon_3(k=0)=1-\frac{2}{3}\frac{\omega_p^2}{\omega^2}.
\end{equation}
The imaginary part ${\rm Im}(\tilde{\epsilon}_3)={\rm Im}(\epsilon_3)$, at $k\rightarrow 0$, is dominated by the $n=1$ Bessel function, and the integral in Eq.~(\ref{eq:eps}) is calculated by taking the value at the pole $\gamma =\frac{\Omega_B}{\omega}$, giving the instability criterion Eq.~(\ref{eq:crit}).

\subsection{Average maser growth rate}
\label{sec:growth}
Here we compute the polarization-averaged growth rate in the limit $\gamma _c\rightarrow \infty$, $B\rightarrow 0$ (the limits are taken in this order, $\omega_p$ is fixed).

First, write the exact dispersion law Eq.~(\ref{eq:edisp}) as
\begin{equation}\label{eq:edispm}
\tilde{\epsilon}_2\big(\tilde{\epsilon}_1\tilde{\epsilon}_3-\tilde{g}_2^2\big)=\tilde{\epsilon}_3g_1^2-2\tilde{g}_2g_1g_3+\tilde{\epsilon}_1g_3^2,
\end{equation}
and compute all of its terms at zero magnetization using Eq.~(\ref{eq:epsd}):
\begin{equation}
\begin{array}{c}
\tilde{\epsilon}_2=\tilde{\epsilon}_\perp,~~ \tilde{\epsilon}_1\tilde{\epsilon}_3-\tilde{g}_2^2=\epsilon_\parallel\tilde{\epsilon}_\perp,~g_1=g_3=0,
\\
\\
\tilde{\epsilon}_\perp\equiv \epsilon_\perp-\frac{k^2}{\omega^2}.
\end{array}
\end{equation}
The approximate dispersion law then reads
\begin{equation}
\epsilon_\parallel\tilde{\epsilon}_\perp=0,
\end{equation}
and gives
\begin{equation}
\omega^2=k^2+\omega_p^2.
\end{equation}

When the magnetic field is turned on, the two terms in the l.h.s. of Eq.~(\ref{eq:edispm}) receive corrections:
\begin{equation}
\delta\tilde{\epsilon}_2\equiv \delta_1,~~ \delta(\tilde{\epsilon}_1\tilde{\epsilon}_3-\tilde{g}_2^2)\equiv \delta_2,
\end{equation}
while the r.h.s. of Eq.~(\ref{eq:edispm}) is
\begin{equation}
\tilde{\epsilon}_1g_3^2+\tilde{\epsilon}_3g_1^2-2\tilde{g}_2g_1g_3\approx g^2,~~g\equiv \hat{k}_zg_1-\hat{k}_xg_3,
\end{equation}
where we have approximated
\begin{equation}
\tilde{\epsilon}_1\approx \hat{k}_x^2,~~~ \tilde{\epsilon}_3\approx \hat{k}_z^2,~~~\tilde{g}_2\approx \hat{k}_x\hat{k}_z,
\end{equation}
which is valid at $k\gg \omega_p$.

Now the dispersion law reads
\begin{equation}\label{eq:dlpr}
(\tilde{\epsilon}_\perp+\delta_1)(\epsilon_\parallel\tilde{\epsilon}_\perp+\delta_2)=g^2.
\end{equation}
To leading order, for $\omega=\sqrt{k^2+\omega_p^2}+\delta\omega$, with $k\gg \omega_p$, we can write
\begin{equation}
\tilde{\epsilon}_\perp=1-\frac{k^2+\omega_p^2}{\omega^2}\approx \frac{2}{\omega}\delta\omega,~~~\epsilon_\parallel\approx 1,
\end{equation}
giving the final form of the dispersion law:
\begin{equation}\label{eq:fdisp}
\big(\frac{2}{\omega}\delta\omega+\delta_1\big)\big(\frac{2}{\omega}\delta\omega+\delta_2\big)=g^2.
\end{equation}

The solution of Eq.~(\ref{eq:fdisp}) is
\begin{equation}\label{eq:fdisps}
\delta \omega=\frac{\omega}{4}\big\{-(\delta_1+\delta_2)\pm \sqrt{(\delta_1-\delta_2)^2+4g^2}\big\}.
\end{equation}
The polarization-averaged frequency correction is
\begin{equation}
\delta \omega=-\frac{\omega}{4}(\delta_1+\delta_2),
\end{equation}
and the polarization-averaged growth rate is
\begin{equation}
\omega_I=-\frac{\omega}{4}({\rm Im}\delta_1+{\rm Im}\delta_2).
\end{equation}
Here
\begin{equation}
{\rm Im}\delta_1={\rm Im}\epsilon_2,
\end{equation}
and a calculation similar to that of \S~\ref{sec:prop} gives the imaginary part ${\rm Im}\delta_1$ as in Eq.~(\ref{eq:ime3}), with a replacement
\begin{equation}
\Omega_B\rightarrow\Omega_B \sin \theta, ~~~\sin \theta\equiv\hat{k}_x.
\end{equation}

For the second imaginary part
\begin{equation}
\begin{array}{c}
{\rm Im}\delta_2={\rm Im}(\tilde{\epsilon}_1\tilde{\epsilon}_3-\tilde{g}_2^2)\approx \tilde{\epsilon}_1{\rm Im}\epsilon_3+\tilde{\epsilon}_3{\rm Im}\epsilon_1-2\tilde{g}_2{\rm Im}g_2
\\
\\

\approx {\rm Im}(\hat{k}_x^2\epsilon_3+\hat{k}_z^2\epsilon_1-2\hat{k}_x\hat{k}_zg_2),
\end{array}
\end{equation}
we must recall that the corresponding $\Pi$-components are
\begin{equation}
\Pi^{(n)}_{33}=v_z^2J_n^2,~~\Pi^{(n)}_{11}=v_\perp^2\frac{n^2}{b^2}J_n^2,~~\Pi^{(n)}_{13}=v_\perp v_z\frac{n}{b}J_n^2.
\end{equation}
When imaginary parts of the permittivity are calculated, as in \S~\ref{sec:prop}, Bessel functions nearly vanish unless $b\approx n$. To sufficient accuracy, we can put $b=n$ in the multiplicative factors of the $\Pi$-components (but not in the argument of $J_n$). Then
\begin{equation}
{\rm Im}\delta_2\propto \hat{k}_x^2v_z^2+\hat{k}_z^2v_\perp^2-2\hat{k}_x\hat{k}_zv_\perp v_z\propto \sin ^2(\chi-\theta),
\end{equation}
where $\theta$, $\chi$ are the propagation and pitch angles ($\hat{k}_x\equiv \sin \theta$, $v_\perp \equiv \sin \chi$). Now a calculation similar to that of \S~\ref{sec:prop} gives the imaginary part ${\rm Im}\delta_2$ as in Eq.~(\ref{eq:ime3}), with replacements
\begin{equation}
\Omega_B\rightarrow\Omega_B \sin \theta,~~~ f_\perp\rightarrow f_\parallel.
\end{equation}

The final result is the polarization-averaged growth rate for propagation at an arbitrary angle $\theta$ at frequency $\omega$:
\begin{equation}\label{eq:pola}
\begin{array}{c}
\omega _I\approx \frac{1}{8\sqrt{3}}\frac{\omega _p^2\Omega _p^2}{\omega^3}\int d^3p\frac{dF}{d\gamma}f_{av} (x),
\\

\\
x\equiv\frac{2}{3}\frac{\gamma \omega_p^3}{\Omega_B\sin \theta\omega^2},
\\

\\
f_{av}(x)\equiv \int\limits_x^\infty dyK_{5/3}(y).
\end{array}
\end{equation}
The maser growth rate in a tangled field is obtained by averaging over propagation directions:
\begin{equation}
\omega _I\approx \frac{1}{2}\int d\theta \sin \theta ~\omega _I(\theta).
\end{equation}

\subsection{Inclined propagation: polarization and growth rates.}
\label{sec:incl}
In \S\ref{sec:modes} we numerically compute the growth rates and polarizations for inclined propagation at $\xi_B\sim 0.001$, $\gamma _c\sim 1000$. Unexpectedly, we find that our numerical results are clearly at odds with the standard synchrotron maser theory of \cite{Ginzburg89, SW02}, although the polarization-averaged growth rate does approximately agree with the standard theory. Here we explain the reason why.

The standard theory says that if the growth rate is much smaller than the frequency shift between the ordinary and extraordinary modes, the modes are circularly polarized and have the same growth rates. If, on the other hand, the growth rate is much larger than the frequency shift between the ordinary and extraordinary modes, the modes are linearly polarized and have different growth rates.

At first site, our theory of \S\ref{sec:growth} does recover this standard result. Consider the dispersion law Eq.(\ref{eq:fdisps}). The real frequency shift between the ordinary and extraordinary modes is given by the $g$ term, which is, using the cold plasma permittivity, but with Lorentz factors included into $\omega _p$ and $\omega_B$,
\begin{equation}
g\sim \frac{\omega _p^2\omega_B}{\omega ^3}.
\end{equation}
As seen from Fig.1, the maser operates at
\begin{equation}
\omega \sim \Big(\frac{\omega_p}{\omega_B}\Big)^\frac{1}{2}\omega_p
\end{equation}
with the growth rate
\begin{equation}
\omega_I \sim \Big(\frac{\omega_B}{\omega_p}\Big)^\frac{3}{2}\omega_p.
\end{equation}
We thus have
\begin{equation}
g\sim \Big(\frac{\omega_B}{\omega_p}\Big)^\frac{5}{2}\ll \frac{\omega _I}{\omega} \sim \Big(\frac{\omega_B}{\omega_p}\Big)^2.
\end{equation}
In the limit $\xi_B\rightarrow0$, the growth rate is much larger than the frequency shift, and the modes must be linearly polarized with different growth rates, given by $f_\perp$ and $f_\parallel$ for the polarization perpendicular and parallel to the magnetic field projection ``on the sky''. And indeed, for $\delta \gg g$, the dispersion law, Eq.(\ref{eq:fdisps}), gives different frequencies, $\delta \omega = -\omega \delta _{1,2}/2$. The polarization, assuming negligible $g_{1,3}$ is clearly linear, either perpendicular (that is along $\hat{y}$) or parallel to the magnetic field projection on the sky.

However, in practice, for small but realistically finite $\xi _B$, the $g$ term turns out to be much larger than $\frac{\omega _I}{\omega}$. This is because, as one can see from Fig.~1, for the fastest growing mode
\begin{equation}
g\approx 5\Big(\frac{\omega_B}{\omega_p}\Big)^\frac{5}{2}> \frac{\omega _I}{\omega} \sim 0.1\Big(\frac{\omega_B}{\omega_p}\Big)^2\quad {\rm for}\quad \xi_B\gtrsim 3\times 10^{-8}.
\end{equation}
This, by itself, is not a problem for the standard maser theory -- just use the opposite limiting case -- circular polarizations with the same growth rate. And indeed, for $\delta \ll g$, assuming that $\epsilon_{1,2,3}$ and $g_2$ can be computed using the zero magnetization formulas, Eq.(\ref{eq:epsd}), while $g_{1,3}$ are arbitrary, we get, after some manipulations, the eigenmode equation
\begin{equation}\label{eq:cip}
\left(
\begin{array}{ccc}
ig_1 & \tilde{\epsilon}_\perp \hat{k}_z & \epsilon_\parallel\hat{k}_x \\

\\

ig_2 & -\tilde{\epsilon}_\perp \hat{k}_x & \epsilon_\parallel\hat{k}_z \\

\\

{\epsilon}_\perp&  -ig & -ih
\end{array}
\right)\left(
\begin{array}{c}
E_\perp \\

\\

E_\parallel \\

\\

E_l
\end{array}
\right)=0,
\end{equation}
where
\begin{equation}
E_\perp \equiv E_y,~~E_\parallel \equiv \hat{k}_zE_x-\hat{k}_xE_z, ~~E_l \equiv \hat{k}_xE_x+\hat{k}_zE_z,
\end{equation}
are the electric field components perpendicular and parallel to the magnetic field projection on the sky, and the longitudinal component parallel to $\hat{k}$. Also,
\begin{equation}
h\equiv \hat{k}_xg_1+\hat{k}_zg_3,~~~g \equiv \hat{k}_zg_1-\hat{k}_xg_3,
\end{equation}
with $g$ as before in \S\ref{sec:growth}.

We now get the dispersion law
\begin{equation}
\epsilon_\parallel\tilde{\epsilon}_\perp^2=\epsilon_\parallel g^2+\tilde{\epsilon}_\perp h^2.
\end{equation}
For polarization, from the first two equations of (\ref{eq:cip}),
\begin{equation}
igE_\perp+\tilde{\epsilon}_\perp E_\parallel=0.
\end{equation}
For electromagnetic modes, we have $\tilde{\epsilon}_\perp\approx 0$, and our dispersion law agrees with Eq.~(\ref{eq:dlpr}), recalling that $\epsilon_\parallel=1$ has been used in Eq.~(\ref{eq:dlpr}). Now, with $\tilde{\epsilon}_\perp^2=g^2$, the polarization is seen to be circular. Also, Eq.~(\ref{eq:fdisps}) predicts the same growth rates for the two modes, assuming that $g$ is real.

However, we will see from numerical calculations that $g$ is not exactly real. It is true that $|{\rm Re}~g|~\gg ~|{\rm Im} ~g|$, but the latter happens to be comparable to $|{\rm Im} ~\delta _{1,2}|$. As a result, we get nearly circularly polarized modes, but with very different growth rates. This is not accounted for by the standard maser theory and we must explain it.

To this end, we write down the imaginary parts of $\delta _{1,2}$ and $g$, deliberately using the same notations as \cite{Ginzburg89}, with the Airy function and its derivative replaced by the modified Bessel functions ${\rm K}_{1/3}$, ${\rm K}_{2/3}$, and with $\psi\equiv \chi -\theta$ (recall that $\chi$ is the electron pitch angle and $\theta$ is the wave propagation angle). ${\rm Im} ~g$ is calculated similar to ${\rm Im} ~\delta _2$, as described in \S\ref{sec:growth}. Replacing, as in  Eq.~(\ref{eq:ime2}), $\psi \rightarrow \frac{\omega_p}{\omega}\psi$, we have
\begin{equation}\label{eq:syncp}
\begin{array}{c}
{\rm Im}~\delta_1\propto \int d\psi ~(1+\psi^2)^2~{\rm K}_{2/3}^2\Big(\frac{1}{2}x(1+\psi^2)^{3/2}\Big),
\\

\\
{\rm Im}~\delta_2\propto \int d\psi ~\psi^2~(1+\psi^2)~{\rm K}_{1/3}^2\Big(\frac{1}{2}x(1+\psi^2)^{3/2}\Big),
\\

\\
{\rm Im}~g\propto \int d\psi ~\psi ~(1+\psi^2)^{3/2}~{\rm K}_{1/3}{\rm K}_{2/3}\Big(\frac{1}{2}x(1+\psi^2)^{3/2}\Big),
\\

\\
x\equiv\frac{2}{3}\frac{\gamma \omega_p^3}{\Omega_B\sin \theta\omega^2}.
\\
\end{array}
\end{equation}
These expressions clearly correspond to equations (5.21-23) of \cite{Ginzburg89}, which give synchrotron power with two linear polarizations and a circular polarization. And, as in the theory of synchrotron radiation, the pitch-angle integral vanishes for the circular polarization by parity.

However, the integrals of Eq.~(\ref{eq:syncp}) are only the lowest order approximations in $\frac{\omega_p}{\omega}$. For instance, the integration measure
\begin{equation}
\sin \chi ~d\chi \approx \sin \theta \big( 1+\cot \theta ~\psi)d\psi
\end{equation}
gives an order $\psi$ correction, which drops out of ${\rm Im}~\delta_{1,2}$ by parity, but renders ${\rm Im}~g$ non-zero. As we have discussed above, even at $\xi _B\sim 10^{-3}$, the corrections can be large. Numerical results give
\begin{equation}
|{\rm Re}~g|~\gg ~|{\rm Im} ~g|\sim {\rm Im} ~\delta _1,
\end{equation}
explaining the origin of the unexpected result -- nearly circular polarizations but with different growth rates.

\section{Numerical computations and simulations of the maser instability}
\label{sec:numerics}

Here we do two different things. First we numerically solve the dispersion law  Eq.~(\ref{eq:disp}) and calculate the growth rates and polarizations of the maser, Langmuir, and Bernstein modes. Then, in order to study the nonlinear saturation of the maser instability, we numerically simulate the development of the maser using a simple particle-in-cell code.

\subsection{Maser and other eigenmodes}
\label{sec:modes}

\begin{table}
\begin{center}
\caption{Monoenergetic distribution, perpendicular propagation. \label{tbl:1}}
\begin{tabular}{llrrr}
\tableline\tableline
\\
\\
$\frac{{\rm Re}(\omega)}{\omega_R}$ & $\frac{{\rm Im}(\omega)}{\xi_B\omega_R}$ & $E_\perp$ & $E_\parallel$ & $E_l$
\\
\\
\tableline
\\
 \\
 0.105 & 3.50 & 0.02 & 0 & (-0.96,0.28)
 \\
 0.110 & 2.08 & 0 & 1 & 0
 \\
 0.116 & 3.69 & 0.01 & 0 & (0.37,0.93)
 \\
 0.121 & 2.57 & 0 & 1 & 0
 \\
 0.121 & 3.50 & 1.00 & 0 & (0.06,-0.06)
\\
0.127 & 3.95 & 0.02 & 0 & (0.95,-0.31)
\\
0.133 & 2.96 & 0 & 1 & 0
\\
0.133 & 2.11 & 0.96 & 0 & (-0.02,0.26)
\\
0.139 & 4.12 & 0.04 & 0 & (-0.99,0.16)
\\
0.145 & 3.23 & 0 & 1 & 0
\\
0.145 & 3.44 & 0.93 & 0 & (-0.01,-0.35)
\\
0.151 & 4.40 & 0.06 & 0 & (0.99,-0.08)
\\
0.156 & 3.48 & 0 & 1 & 0
\\
0.156 & 3.58 & 0.93 & 0 & (0.04,0.38)
\\
0.162 & 4.43 & 0.05 & 0 & (-0.98,0.20)
\\
 0.168 & 3.64 & 0 & 1 & 0
 \\
 0.168 & 2.93 & 0.98 & 0 & (-0.06,-0.16)
 \\
 0.174 & 4.51 & 0.02 & 0 & (0.68,0.73)
 \\
 0.179 & 3.78 & 0 & 1 & 0
 \\
 0.179 & 4.59 & 0.99 & 0 & (0.10,-0.02)
 \\
 0.185 & 4.70 & 0.03 & 0 & (0.84,-0.54)
 \\
 0.191 & 3.89 & 0 & 1 & 0
 \\
 0.191 & 3.26 & 0.97 & 0 & (-0.04,0.22)
 \\
 0.197 & 4.68 & 0.07 & 0 & (-0.97,0.22)
\\
\tableline
\tableline

\end{tabular}
\tablecomments{ $\xi_B=10^{-3}$, $k=0.5\omega_R$, $\hat{k}_z=0$, growing eigenmodes with $\frac{{\rm Re}(\omega)}{\omega_R}$ between 0.1 and 0.2. }
\end{center}
\end{table}

A numerical solution of the dispersion law, Eq.~(\ref{eq:disp}), requires a numerical calculation of the permittivity, Eq.~(\ref{eq:eps}). This is a lengthy expression, and the numerical code calculating it must be checked. Our check was to reproduce simpler expressions which must follow from Eq.~(\ref{eq:eps}) in various cases. We checked the cases of a cold plasma, an unmagnetized plasma, and a Maxwellian plasma. In the latter case, as a check, we used the permittivity given in \cite{LP81}, which is a complicated quadrature without explicit Bessel functions. Our numerical permittivity does reproduce these three permittivities.

\begin{table}
\begin{center}
\caption{Smooth hollow distribution, perpendicular propagation. \label{tbl:2}}
\begin{tabular}{lllrrr}
\tableline\tableline
\\
\\
$\frac{k}{\omega_R}$ & $\frac{{\rm Re}(\omega)}{\omega_R}$ & $\frac{{\rm Im}(\omega)}{\xi_B\omega_R}$ & $E_\perp$ & $E_\parallel$ & $E_l$
\\
\\
\tableline
\\
 \\
 0.15 & 0.242 & 0.143 & 0.17 & 0 & (-0.05,-0.98)
 \\
 ~~~ & 0.266 & 0.077 & 0.99 & 0 & (0,0.12)
 \\
 0.20 & 0.266 & 0.366 & 0.15 & 0 & (-0.09,-0.98)
 \\
 ~~~ & 0.300 & 0.148 & 1.00 & 0 & (0,0.07)
 \\
 0.25 & 0.296 & 0.305 & 0.17 & 0 & (-0.07,-0.98)
 \\
 ~~~ & 0.338 & 0.180 & 1.00 & 0 & (0,0.05)
 \\
 ~~~ & 0.337 & 0.012 & 0 & 1 & 0
 \\
 0.30 & 0.333 & 0.042 & 0.24 & 0 & (-0.02,-0.97)
 \\
 ~~~ & 0.378 & 0.217 & 1.00 & 0 & (0,0.04)
 \\
 ~~~ & 0.378 & 0.016 & 0 & 1 & 0
 \\
 0.35 & 0.421 & 0.235 & 1.00 & 0 & (0,0.03)
 \\
 ~~~ & 0.420 & 0.021 & 0 & 1 & 0
 \\
 0.40 & 0.464 & 0.239 & 1.00 & 0 & (0,0.03)
 \\
 ~~~ & 0.464 & 0.028 & 0 & 1 & 0
 \\
 0.45 & 0.509 & 0.196 & 1.00 & 0 & (0,0.03)
 \\
 ~~~ & 0.509 & 0.025 & 0 & 1 & 0
 \\
 0.50 & 0.555 & 0.127 & 1.00 & 0 & (0,0.02)
 \\
 ~~~ & 0.554 & 0.025 & 0 & 1 & 0
 \\
 0.55 & 0.601 & 0.081 & 1.00 & 0 & (0,0.02)
 \\
 ~~~ & 0.600 & 0.032 & 0 & 1 & 0
 \\
 0.60 & 0.647 & 0.020 & 0 & 1 & 0
 \\
 0.65 & 0.694 & 0.014 & 0 & 1 & 0
 \\
\tableline
\tableline

\end{tabular}
\tablecomments{ $\xi_B=10^{-3}$, $\hat{k}_z=0$, all growing eigenmodes for a given $k$.}
\end{center}
\end{table}

We use Eq.~(\ref{eq:eps}) as written, without Landau's analytic continuation into the lower complex $\omega$ half-plane. This means that we calculate only the growing eigenmodes.

We have calculated the growth rates and polarizations for different: (i) shapes of the distribution function, (ii) characteristic Lorentz factors $\gamma_c$, (iii) magnetizations $\xi_B$, (iv) wavenumbers $k$, (v) propagation directions $\hat{k}$. To concisely illustrate our findings we fix, once and for all, two of these five variables:

\begin{enumerate}

\item {\bf Lorentz factor $\gamma _c=10^3$.} This is motivated by the expected value. Also, this $\gamma$ is large enough, so that various approximations used in the theoretical sections do apply.

\item {\bf Magnetization $\xi_B=10^{-3}$.}  This should be  about the smallest magnetization of the downstream plasma one can expect. Also, this $\xi_B$ is small enough, so that various approximations used in the theoretical sections should roughly apply.

\end{enumerate}

\begin{table}
\begin{center}
\caption{Smooth hollow distribution, inclined propagation. \label{tbl:3}}
\begin{tabular}{lllrrr}
\tableline\tableline
\\
\\
$\frac{k}{\omega_R}$ & $\frac{{\rm Re}(\omega)}{\omega_R}$ & $\frac{{\rm Im}(\omega)}{\xi_B\omega_R}$ & $E_\perp$ & $E_\parallel$ & $E_l$
\\
\\
\tableline
\\
 \\
 0.15 & 0.242 & 0.088 & 0.15 & (0,0.01) & (-0.03,-0.99)
 \\
 ~~~ & 0.267 & 0.055 & 0.74 & (-0.01,-0.67) & (0,0.07)
 \\
 ~~~ & 0.264 & 0.011 & 0.66 & (-0.01,0.75) & (0,0.07)
 \\
 0.20 & 0.266 & 0.277 & 0.13 & (0,0) & (-0.08,-0.99)
 \\
 ~~~ & 0.301 & 0.083 & 0.73 & (-0.02,-0.68) & (0,0.05)
 \\
 ~~~ & 0.299 & 0.022 & 0.68 & (-0.03,0.73) & (0,0.05)
 \\
 0.25 & 0.296 & 0.336 & 0.15 & (0,0) & (-0.08,-0.99)
 \\
 ~~~ & 0.338 & 0.114 & 0.73 & (-0.04,-0.68) & (0,0.03)
 \\
 ~~~ & 0.336 & 0.033 & 0.68 & (-0.04,0.73) & (0,0.03)
 \\
 0.30  & 0.379 & 0.142 & 0.74 & (-0.05,-0.67) & (0,0.03)
 \\
 ~~~ & 0.377 & 0.044 & 0.67 & (-0.06,0.74) & (0,0.02)
 \\
 0.35  & 0.421 & 0.162 & 0.75 & (-0.06,-0.66) & (0,0.02)
 \\
 ~~~ & 0.420 & 0.051 & 0.66 & (-0.07,0.75) & (0,0.02)
 \\
 0.40  & 0.465 & 0.170 & 0.76 & (-0.07,-0.65) & (0,0.02)
 \\
 ~~~ & 0.463 & 0.053 & 0.64 & (-0.08,0.76) & (0,0.02)
 \\
 0.45  & 0.510 & 0.163 & 0.77 & (-0.06,-0.63) & (0,0.02)
 \\
 ~~~ & 0.508 & 0.047 & 0.63 & (-0.08,0.78) & (0,0.01)
 \\
 0.50  & 0.555 & 0.142 & 0.79 & (-0.05,-0.61) & (0,0.02)
 \\
 ~~~ & 0.554 & 0.040 & 0.60 & (-0.07,0.79) & (0,0.01)
 \\
 0.55  & 0.601 & 0.111 & 0.81 & (-0.04,-0.59) & (0,0.01)
 \\
 ~~~ & 0.600 & 0.030 & 0.59 & (-0.05,0.81) & (0,0.01)
 \\
 0.60  & 0.648 & 0.072 & 0.82 & (-0.02,-0.57) & (0,0.01)
 \\
 ~~~ & 0.647 & 0.018 & 0.57 & (-0.02,0.82) & (0,0.01)
 \\
 0.65  & 0.695 & 0.035 & 0.83 & (0.01,-0.56) & (0,0.01)
 \\
 ~~~ & 0.694 & 0.004 & 0.55 & (0.02,0.84) & (0,0.01)

  \\
\tableline
\tableline

\end{tabular}
\tablecomments{ $\xi_B=10^{-3}$, $\hat{k}_z=0.5$, all growing eigenmodes for a given $k$.}
\end{center}
\end{table}

To show the polarization of the eigenmodes, we decompose the (complex) electric field vector into two transverse components: $E_\perp$ (perpendicular to $\hat{k}$ and $\hat{B}$, taken to be real), $E_\parallel$ (perpendicular to $\hat{k}$ and $E_\perp$), and a longitudinal component: $E_l$ (along $\hat{k}$), and normalize by $E_\perp^2+|E_\parallel|^2+|E_l|^2=1$.

To present the results, we define, following \cite{W17} the modified Razin frequency
\begin{equation}
\omega_R\equiv\Big(\frac{9}{2}\xi_B\Big)^{-1/4}\omega_p.
\end{equation}
As we have shown in \S \ref{sec:mono}, the maser modes should grow the fastest at  $\omega \sim \omega _R$.

The simplest hollow distribution is the monoenergetic distribution considered by \cite{W17}. However, as Table~\ref{tbl:1} illustrates, this distribution actually gives a large number of Bernstein modes, which grow much faster than at the theoretical rates of \S \ref{sec:mono}. Apparently, for a narrow distribution function,  replacing the sum over Larmor harmonics by the integral is incorrect even at small $\xi_B\sim 10^{-3}$.

Now consider a smooth hollow distribution function: $F\propto p^2e^{-2p^2/p_0^2}$. Then, as seen from  Table~ \ref{tbl:2}, the Bernstein modes disappear; only the theoretically predicted modes exist: the EM modes (mostly transverse, what we call the maser modes) and the Langmuir mode (mostly longitudinal). The growth rates, especially the maser growth rates,  are in reasonable agreement with the theoretical predictions.

Table~\ref{tbl:3} lists unstable modes propagating at the median (by solid angle) inclination $\hat{k}_z=0.5$, for the same smooth hollow distribution function. We see that the growth rates are close to those for the perpendicular propagation. The polarization-averaged growth rate of the maser modes given by Eq.~(\ref{eq:pola}) agrees with the numerical results to about 30\%.

\subsection{Maser and other modes: non-linear saturation}

Linear instabilities, which we apparently do understand, characterize only the initial stage of the plasma dynamics. If the shock quickly prepares the initially hollow distribution, the growing modes will be generated from the background noise. But the long-term evolution necessarily involves non-linear effects, and these are best studied by direct numerical simulations, the so-called particle-in-cell (PIC) simulations. In fact, until a full three-dimensional PIC simulation is  done, we cannot be sure that some non-linear interactions don't completely change the initially linear evolution, so much so as to make our linear analysis entirely misleading.

\begin{figure}
 \includegraphics[width=0.5\textwidth]{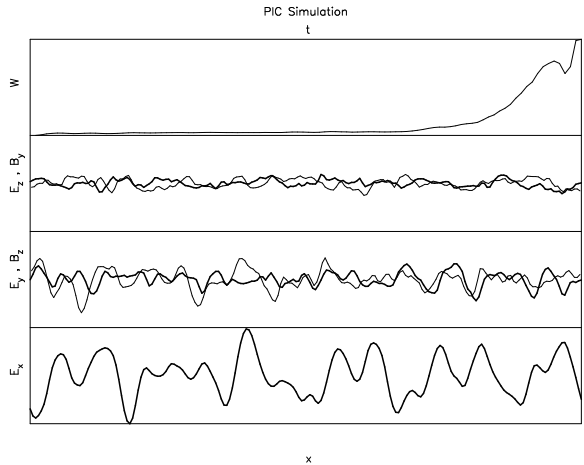}
\caption{Upper curve -- the energy of the EM field perturbation, $W$, as a function of time; maximum time $=50 \omega_p^{-1}$, maximum  $W=0.043$ of the unperturbed field energy. Lower curves --- electric fields (thick) and magnetic fields (thin, uniform field subtracted) as a function of the coordinate. The box size $= 40 \omega_p^{-1}$. }
\label{fig:pic1}
\end{figure}

We have run such a fully 3D simulation, and got results similar to the 1x3v (one spatial dimension, three velocity dimensions) PIC simulation  that we describe below. The reason we choose to describe only this toy 1x3v simulation is that the maser instabilities are slow, and we were able to run the fully 3D code only at low resolution, and we can not be entirely sure that we got it right.

The 1x3v simulation is as follows. Consider an EM field which depends only on $t$ and $x$: $E_x(t,x)$,  $E_y(t,x)$, $E_z(t,x)$, $B_x={\rm const}$, $B_y(t,x)$, $B_z(t,x)$ (our x-grid has 200 points). Add a large number of charged particles (we had $2\times 10^6$ particles), each of which is characterized by the (time-dependent) x-coordinate, and three velocity components. Let the charges interact with the fields by the Maxwell-Lorentz equations. Start with a constant magnetic field, zero electric field, and a particle distribution which is uniform in space, and isotropic and hollow in the velocity space. The code is the (straightforwardly modified) TRISTAN, \cite{Buneman93}.

\begin{figure}
 \includegraphics[width=0.5\textwidth]{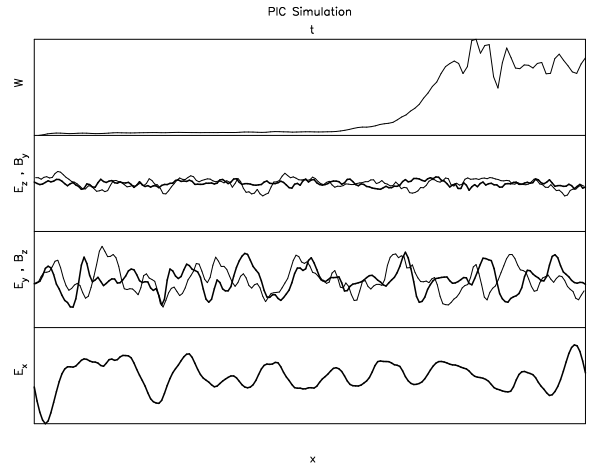}
\caption{Same as Fig.~(\ref{fig:pic1}), but with maximum time $=62.5 \omega_p^{-1}$. }
\label{fig:pic2}
\end{figure}

We show the simulation results for perpendicular propagation -- the initial uniform magnetic field is along the z-axis. We have $\xi_B=0.01$ in an electron-positron plasma. The initial distribution function is isotropic and monoenergetic with $\gamma=3$. As seen in Fig.~(\ref{fig:pic1}), Langmuir modes grow the fastest, saturating when the EM field (including electrostatic) perturbation energy reaches few percent of the uniform magnetic field energy. Note that, as expected, the perpendicular polarized transverse modes are stronger than the parallel polarized transverse modes.

As seen in Fig.~(\ref{fig:pic2}), upon saturation of the Langmuir modes, the perpendicular polarized transverse modes keep growing, and become comparable in energy to the Langmuir modes, while the parallel polarized transverse modes remain weak.

\section{Discussion}
\label{sec:discussion}

We have derived exact (numeric) solutions of the dispersion equation for a weakly magnetized (magnetic field to particle energy density ratio $\xi_B\ll1$) highly relativistic (electron Lorentz factor $\gamma_c\gg1$) plasma with an isotropic particle distribution function, starting from the exact dielectric permittivity tensor. We have shown that unstable maser modes exist for a "hollow particle distribution" provided that $\gamma_c^2\xi_B>1$. By "hollow" we mean a distribution with particle density (in momentum space), $F$, that increases with momentum (i.e. $dF/d\gamma>0$, or $dn/d\gamma\propto \gamma^2 dF/d\gamma$ rising faster than $\gamma^2$). A sufficient condition for maser instability is given by Eq.~(\ref{eq:crit}).

The maser modes grow fastest (see fig.~\ref{fig:emgr}, \S~\ref{sec:growth}) near the modified Razin frequency, $\omega_R\approx\xi_B^{-1/4}\omega_p$, at a rate $\omega_{IR}\approx\xi_B^{3/4}\omega_p$ (see Eqs.~\ref{eq:wR},~\ref{eq:growth} for exact results; $\omega_p$ is the plasma frequency, $\omega_p^2\approx4\pi ne^2/(\gamma_c m)$ see Eqs.~\ref{eq:Om} and~\ref{eq:omp0}). These results are similar to those obtained using the standard maser theory. Moreover, for mono-energetic electrons and perpendicular propagation, i.e. for wave vectors perpendicular to the magnetic field direction, the analytic expression derived for the growth rates from the exact dispersion relation, eq.~(\ref{eq:monem}), is identical to that obtained by the standard maser theory, eqs.~(34-37) of \citet{W17}. This exact agreement supports the validity of both current and earlier analytic work, as well as the validity of our current numerical results, which agree with the analytic results in the appropriate limits.

A significant deviation of our exact results from those of the approximate standard theory is related to the polarization of the growing modes. For inclined propagation, in a direction not perpendicular to the field, and a small but realistic field, $\xi_B>3\times10^{-8}$, the standard maser theory predicts circularly polarized modes with similar growth rates (see \S~\ref{sec:results-pol}, Eq.~\ref{eq:xiBmax}). Our exacts results indeed show that the growing modes are nearly circularly polarized for $\xi_B=10^{-3}$. However, the growth rates of the two modes are not similar. We showed in \S~\ref{sec:incl} that this deviation is due to circularly polarized synchrotron emission, which is neglected in the standard theory.

Our analysis, based on a direct solution for the zeros of the exact dispersion equation, enabled us to compare the growth rate of the maser mode to those of non-electromagnetic modes. We find that the growth rate of Langmuir (electro-static) waves is larger than that of the maser mode. The ratio of Langmuir and maser modes growth rates is larger for smaller $\xi_B$ ($\propto \xi_B^{-5/12}$). For $\xi_B=10^{-3}$, which is the smallest value expected in the downstream of a collisionless shock, the growth rate ratio is only a few. Thus, the maser mode growth is not expected to be suppressed. Moreover, we have investigated numerically in \S~\ref{sec:numerics} the nonlinear evolution of the instability for $\xi_B=10^{-3}$, and found that the growth of both the maser and Langmuir modes saturates at a similar energy density of the waves, which is a few percent of the initial (uniform) magnetic field energy density (the Langmuir modes reach saturation faster).

Collisionless shock waves are responsible for a wide range of high energy astrophysical phenomena. The analysis of this paper implies that if such shocks produce hollow distribution functions, then (i) the maser instability may play a significant role in the thermalization of the particle distribution, and (ii) a significant fraction of the shock energy may be radiated away as electromagnetic waves at frequencies somewhat higher than the plasma frequency. Such conversion is difficult to achieve in general, and it may lead to coherent radio emission with very large brightness temperatures. In \S~\ref{sec:relevance} we provide some order of magnitude estimates for the conditions under which bright coherent maser radio emission may be expected. These estimates may be used as a guidance when studying particular astrophysical sources/phenomena. Two shock configurations were considered: a shock propagating into a plasma at rest (i.e. not expanding relativistically, \S~\ref{sec:rest}), and a shock propagating within a relativistically expanding plasma (\S~\ref{sec:jets}).

We find that coherent maser emission may be produced by relativistic winds driven by black-holes over a wide range of masses, provided that the energy flux is dominated by kinetic energy at the dissipation region. For example, minute time scale $L\sim10^{41}{\rm erg/s}$ radio bursts may be produced by $\sim10^6M_\odot$ black holes. Possible sites for the generation of observable maser radio emission may be produced by a shock propagating into a non-relativistically-expanding plasma are the hot corona regions around accretion disks near compact objects. The plasma density may be large enough in such systems for a (mildly) relativistic shock to produce radio maser emission, and the systems may be compact and hot enough to avoid free-free absorption.

\acknowledgements  This research was partially supported by ISF, IMOS and Minerva grants. AG acknowledges the hospitality of Slava Mukhanov of LMU, where this work has been completed.

\vskip 1cm

\bibliographystyle{hapj}

\end{document}